# NONLINEAR SELF-ORGANIZATION DYNAMICS OF A METABOLIC PROCESS OF THE KREBS CYCLE

V. I. GRYTSAY[a,*], I. V. MUSATENKO[b]

[a]Bogolyubov Institute for Theoretical Physics, 14b, Metrolohichna Str., Kyiv 03680, Ukraine;
E-mail: vgrytsay@bitp.kiev.ua
[b] Taras Shevchenko National University of Kyiv, Faculty of Cybernetics, Department of Computational Mathematics, 64, Volodymyrska Str., Kyiv, Ukraine
E-mail: ivmusatenko@gmail.com

The present work continues studies of the mathematical model of a metabolic process of the Krebs cycle. We study the dependence of its cyclicity on the cell respiration intensity determined by the formation level of carbon dioxide. We constructed the phase-parametric characteristic of the consumption of a substrate by a cell depending on the intensity of the metabolic process of formation of the final product of the oxidation. The scenarios of all possible oscillatory modes of the system are constructed and studied. The bifurcations with period doubling and with formation of chaotic modes are found. Their attractors are constructed. The full spectra of indices and divergencies for the obtained modes, the values of KS-entropies, horizons of predictability, and Lyapunov dimensions of strange attractors are calculated. Some conclusions about the structural-functional connections of the cycle of tricarboxylic acids and their influence on the stability of the metabolic process in a cell are presented.

*K e y w o r d s :* Krebs cycle, metabolic process, self-organization, strange attractor, bifurcation, Feigenbaum scenario.

One of the possible problems of synergetics is the study of the internal dynamics of metabolic processes in cells. Its solution allows one to find the structural-functional connections defining the self-organization of these processes and to answer the question how the catalyzed enzymatic reactions create the internal space-time ordering of the cell life.

The most general metabolic process in cells is the cycle of tricarboxylic acids [1]. This is the key stage of the respiration of all cells. In its course, the di- and tricarbon compounds, which are formed as intermediate products in the transformation of carbohydrates, fats, and proteins, are transformed up to $CO_2$. In this case, the released hydrogen is oxidized further up to water, by taking the direct participation in the synthesis of ATP, being the universal energy source.

Studies of the functioning of the Krebs cycle were carried out both experimentally and theretically in [2-10].

In the study of the given process, we use the mathematical model of the growh of cells *Candida utilis* on ethanol, which was developed by Professor V.P. Gachok [11, 12]. The analogous modeling of a growth of cells was performed independently by J. Monod, V.S. Podgorskii, L.N. Drozdov-Tikhomirov, N.T. Rakhimova, G.Yu. Riznichenko, and other researchers [13-17].

With the help of this model, the unstable modes in the cultivation of cells observed in experiments were considered. The kinetic curves of the chaotic dynamics obtained with the help of computational experiments were in agreement with experimental data [18].

Then the given model was modified and refined in [19] due to the account for the influence of the $CO_2$ level on the respiration intensity. With the help of the model, the structural-functional connections of the metabolic process in a cell, which cause the appearance of complicated oscillations in the metabolic process, were investigated. It was concluded that the given oscillations arise on the level of redox reactions of the Krebs cycle, reflect the cyclicity of the process, and characterize the self-organization in a cell. The fractality of the dynamics of oscillations of the Krebs cycle was studied as well.

The analogous oscillatory modes were observed in the processes of photosynthesis and glycolysis, variations of the calcium concentration in a cell, oscillations in heart muscle, and other biochemical processes [20-24].

## MATHEMATICAL MODEL

The general scheme of the process is presented in Fig. 1. According to it with regard for the mass balance, we have constructed the mathematical model given by Eqs. (1) - (19).

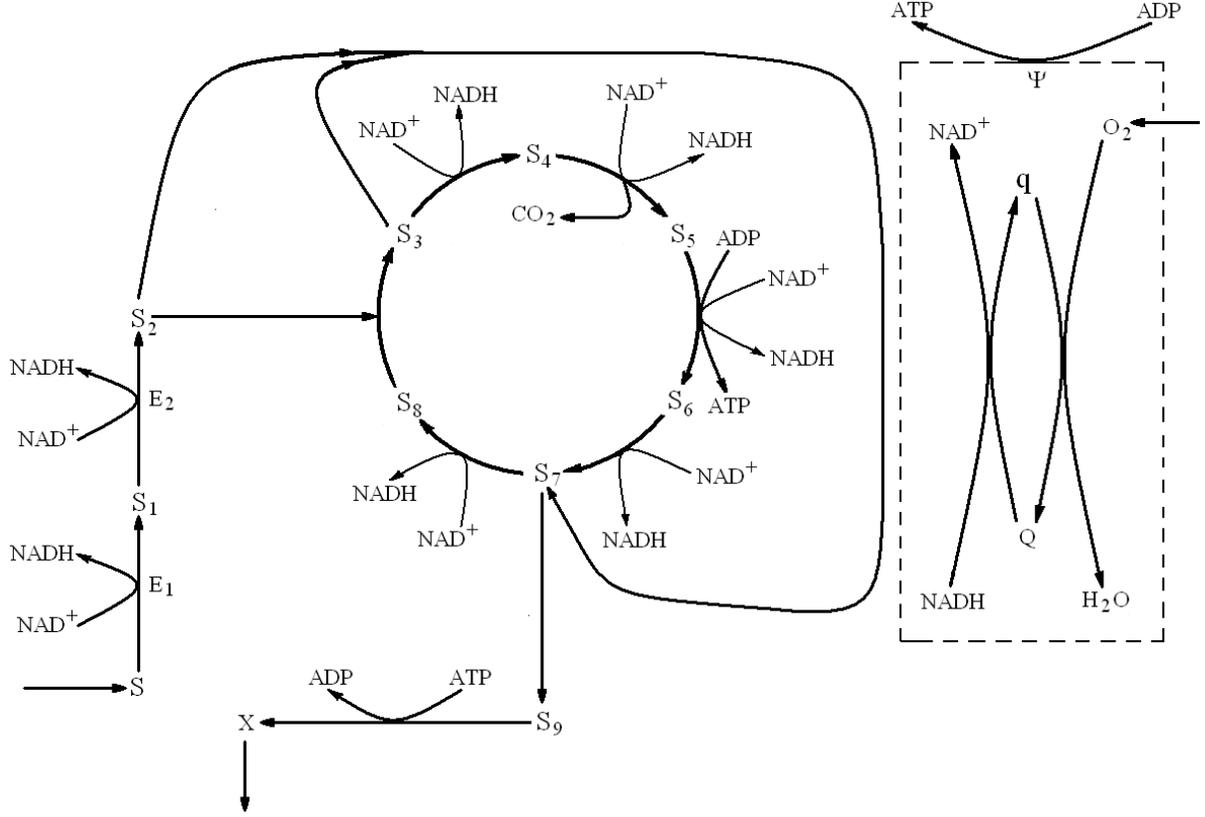

Fig. 1. General scheme of the metabolic process of growth of cells *Candida utilis* on ethanol.

$$\frac{dS}{dt} = S_0 \frac{K}{K+S+\gamma\psi} - k_1 V(E_1) \frac{N}{K_1+N} V(S) - \alpha_1 S, \tag{1}$$

$$\frac{dS_1}{dt} = k_1 V(E_1) \frac{N}{K_1+N} V(S) - k_2 V(E_2) \frac{N}{K_1+N} V(S_1), \tag{2}$$

$$\frac{dS_2}{dt} = k_2 V(E_2) \frac{N}{K_1+N} V(S_1) - k_3 V(S_2^2) V(S_3) - k_4 V(S_2) V(S_8), \tag{3}$$

$$\frac{dS_3}{dt} = k_4 V(S_2) V(S_8) - k_5 V(N^2) V(S_3^2) - k_3 V(S_2^2) V(S_3), \tag{4}$$

$$\frac{dS_4}{dt} = k_5 V(N^2) V(S_3^2) - k_7 V(N) V(S_4) - k_8 V(N) V(S_4), \tag{5}$$

$$\frac{dS_5}{dt} = k_7 V(N) V(S_4) - 2k_9 V(L_1 - T) V(S_5), \tag{6}$$

$$\frac{dS_6}{dt} = 2k_9 V(L_1 - T)V(S_5) - k_{10} V(N) \frac{S_6^2}{S_6^2 + 1 + M_1 S_8}, \tag{7}$$

$$\frac{dS_7}{dt} = k_{10} V(N) \frac{S_6^2}{S_6^2 + 1 + M_1 S_8} - k_{11} V(N) V(S_7) - k_{12} \frac{S_7^2}{S_7^2 + 1 + M_2 S_9} V(\psi^2) + k_3 V(S_2^2) V(S_3), \tag{8}$$

$$\frac{dS_8}{dt} = k_{11} V(N) V(S_7) - k_4 V(S_2) V(S_8) + k_6 V(T^2) \frac{S^2}{S^2 + \beta_1} \cdot \frac{N_1}{N_1 + (S_5 + S_7)^2}, \tag{9}$$

$$\frac{dS_9}{dt} = k_{12} \frac{S_7^2}{S_7^2 + 1 + M_2 S_9} V(\psi^2) - k_{14} \frac{XTS_9}{(\mu_1 + T)[(\mu_2 + S_9 + X + M_3(1 + \mu_3\psi)]S}, \tag{10}$$

$$\frac{dX}{dt} = k_{14} \frac{XTS_9}{(\mu_1 + T)[(\mu_2 + S_9 + X + M_3(1 + \mu_3\psi)]S} - \alpha_2 X, \tag{11}$$

$$\frac{dQ}{dt} = -k_{15} V(Q) V(L_2 - N) + 4k_{16} V(L_3 - Q) V(O_2) \frac{1}{1 + \gamma_1 \psi^2}, \tag{12}$$

$$\frac{dO_2}{dt} = O_{2_0} \frac{K_2}{K_2 + O_2} - k_{16}(L_3 - Q) V(O_2) \frac{1}{1 + \gamma_1 \psi} - k_8 V(N) V(S_4) - \alpha_3 O_2, \tag{13}$$

$$\frac{dN}{dt} = -k_7 V(N) V(S_4) - k_{10} V(N) \frac{S_6^2}{S_6^2 + 1 + M_1 S_8} - k_{11} V(N) V(S_7) - k_5 V(N^2) V(S_3^2) + $$
$$+ k_{15} V(Q) V(L_2 - N) - k_2 V(E_2) \frac{N}{K_1 + N} V(S_1) - k_1 V(E_1) \frac{N}{K_1 + N} V(S), \tag{14}$$

$$\frac{dT}{dt} = k_{17} V(L_1 - T) V(\psi^2) + k_9 V(L - T) V(S_3) - \alpha_4 T - $$
$$- k_{18} k_6 V(T^2) \frac{S^2}{S^2 + \beta_1} \cdot \frac{N_1}{N_1 + (S_5 + S_7)^2} - k_{19} k_{14} \frac{XTS_9}{(\mu_1 + T)[\mu_2 + S_9 + X + M_3(1 + \mu_3\psi)S]}, \tag{15}$$

$$\frac{d\psi}{dt} = 4k_{15} V(Q) V(L_2 - N) + 4k_{17} V((L_1 - T) V(\psi^2) - 2k_{12} \frac{S_7^2}{S_7^2 + 1 + M_2 S_9} V(\psi^2) - \alpha\psi, \tag{16}$$

$$\frac{dE_1}{dt} = E_{1_0} \frac{S^2}{\beta_2 + S^2} \frac{N_2}{N_2 + S_1} - n_1 V(E_1) \frac{N}{K_1 + N} V(S) - \alpha_5 E_1, \tag{17}$$

$$\frac{dE_2}{dt} = E_{2_0} \frac{S_1^2}{\beta_3 + S_1^2} \frac{N_3}{N_3 + S_2} - n_2 V(E_2) \frac{N}{K_1 + N} V(S_1) - \alpha_6 E_2, \tag{18}$$

$$\frac{dC}{dt} = k_8 V(N) V(S_4) - \alpha_7 C. \tag{19}$$

where $V(X) = X/(1 + X)$ is the function that describes the adsorption of the enzyme in the region of a local coupling. The variables of the system are dimensionless [18, 19].

The internal parameters of the system are as follows:

$k_1 = 0.3;\ k_2 = 0.3;\ k_3 = 0.2;\ k_4 = 0.6;\ k_5 = 0.16;\ k_6 = 0.7;\ k_7 = 0.08;\ k_8 = 0.022;$
$k_9 = 0.1;\ k_{10} = 0.08;\ k_{11} = 0.08;\ k_{12} = 0.1;\ k_{14} = 0.7;\ k_{15} = 0.27;\ k_{16} = 0.18;$
$k_{17} = 0.14;\ k_{18} = 1;\ k_{19} = 10;\ n_1 = 0.07;\ n_2 = 0.07;\ L = 2;\ L_1 = 2;\ L_2 = 2.5;\ L_3 = 2;$
$K = 2.5;\ K_1 = 0.35;\ K_2 = 2;\ M_1 = 1;\ M_2 = 0.35;\ M_3 = 1;\ N_1 = 0.6;\ N_2 = 0.03;$
$N_3 = 0.01;\ \mu_1 = 1.37;\ \mu_2 = 0.3;\ \mu_3 = 0.01;\ \gamma = 0.7;\ \gamma_1 = 0.7;\ \beta_1 = 0.5;\ \beta_2 = 0.4;$
$\beta_3 = 0.4;\ E_{1_0} = 2;\ E_{2_0} = 2.$

The external parameters determining the flow-type conditions are chosen as
$S_0 = 0.05055;\ O_{2_0} = 0.06;\ \alpha = 0.002;\ \alpha_1 = 0.02;\ \alpha_2 = 0.004;\ \alpha_3 = 0.01;$
$\alpha_4 = 0.01;\ \alpha_5 = 0.01;\ \alpha_6 = 0.01;\ \alpha_7 = 0.0001.$

The model covers the processes of substrate-enzymatic oxidation of ethanol to acetate, cycle involving tri- and dicarboxylic acids, glyoxylate cycle, and respiratory chain.

The incoming ethanol $S$ is oxidized by the alcohol dehydrogenase enzyme $E_1$ to acetaldehyde $S_1$ (1) and then by the acetal dehydrogenase enzyme $E_2$ to acetate $S_2$ (2), (3). The formed acetate can participate in the cell metabolism and can be exchanged with the environment. The model accounts for this situation by the change of acetate by acetyl-$CoA$. On the first stage of the Krebs cycle due to the citrate synthase reaction, acetyl-$CoA$ jointly with oxalacetate $S_8$ formed in the Krebs cycle create citrate $S_3$ (4). Then substances $S_4$ - $S_8$ are created successively on stages (5)-(9). In the model, the Krebs cycle is represented by only those substrates that participate in the reduction of $NADH$ and the phosphorylation $ADT \to ATP$. Acetyl-$CoA$ passes along the chain to malate represented in the model as intramitochondrial $S_7$ (8) and cytosolic $S_9$ (10) ones. Malate can be also synthesized in another way related to the activity of two enzymes: isocitrate lyase and malate synthetase. The former catalyzes the splitting of isocitrate to succinate, and the latter catalyzes the condensation of acetyl-$CoA$ with glyoxylate and the formation of malate. This glyoxylate-linked way is shown in Fig. 1 as an enzymatic reaction with the consumption of $S_2$ and $S_3$ and the formation of $S_7$. The parameter $k_3$ controls the activity of the активность glyoxylate-linked way (3), (4), (8). The yield of $S_7$ into cytosol is controlled by its concentration, which can increase due to $S_9$, by causing the inhibition of its transport with the participation of protons of mitochondrial membrane.

The formed malate $S_9$ is used by a cell for its growth, namely for the biosynthesis of protein $X$ (11). The energy consumption of the given process is supported by the process $ATP \to ADP$. The presence of ethanol in the external solution causes the "ageing" of external membranes of cells, which leads to the inhibition of this process. The inhibition of the process also happens due to the enhanced level of the kinetic membrane potential $\psi$. The parameter $\mu_0$ is related to the lysis and the washout of cells.

In the model, the respiratory chain of a cell is represented in two forms: oxidized, $Q$, (12) and reduced, $q$, ones. They obey the integral of motion $Q(t) + q(t) = L_3$.

A change of the concentration of oxygen in the respiratory chain is determined by Eq. (13).

The activity of the respiratory chain is affected by the level of $NADH$ (14). Its high concentration leads to the enhanced endogenic respiration in the reducing process in the respiratory chain (parameter $k_{15}$). The accumulation of $NADH$ occurs as a result of the reduction of $NAD^+$ at the transformation of ethanol and in the Krebs cycle. These variables obey the integral of motion $NAD^+(t) + NADH(t) = L_2$.

In the respiratory chain and the Krebs cycle, the substrate-linked phosphorylation of $ADP$ with the formation of $ATP$ (15) is also realized. The energy consumption due to the process $ATP \to ADP$ induces the biosynthesis of components of the Krebs cycle (parameter $k_{18}$) and the growth of cells on the substrate (parameter $k_{19}$). For these variables, the integral of motion $ATP(t) + ADP(t) = L_1$ holds. Thus, the level of $ATP$ produced in the redox processes in the respiratory chain $ADP \to ATP$ determines the intensity of the Krebs cycle and the biosynthesis of protein.

In the respiratory chain, the kinetic membrane potential $\psi$ (16) is created under the running of reducing processes $Q \to q$. It is consumed at the substrate-linked phosphorylation $ADP \to ATP$ in the respiratory chain and the Krebs cycle. Its enhanced level inhibits the biosynthesis of protein and process of reduction of the respiratory chain.

Equations (17) and (18) describe the activity of enzymes $E_1$ and $E_2$, respectively. We consider their biosynthesis ($E_{1_0}$ and $E_{2_0}$), the inactivation in the course of the enzymatic reaction ($n_1$ and $n_2$), and all possible irreversible inactivations ($\alpha_5$ and $\alpha_6$).

Equation (19) is related to the formation of carbon dioxide. Its removal from the solution into the environment ($\alpha_7$) is taken into account. Carbon dioxide is produced in the Krebs cycle (5). In addition, it squeezes out oxygen from the solution (13), by decreasing the activity of the respiratory chain.

The study of solutions of the given mathematical model (1)-(19) was performed with the help of the theory of nonlinear differential equations [25, 26] and the methods of mathematical modeling of biochemical systems applied and developed by the authors in [27-43].

For the numerical solution of this autonomous system of nonlinear differential equations, we used the Runge--Kutta--Merson method. The accuracy of calculatuons is set to be $10^{-8}$. In order to attain the reliability of the study, namely to attain the transition of the system from an initial state to the asymptotic mode characterized by an attractor, we took the calculation time as high as $10^6$. For this time, the trajectory "sticks" on the corresponding attractor.

## THE RESULTS OF STUDIES

For one cycle, there occurs the full oxidation of a molecule of acetyl-$CoA$ up to malate and the formation of a new molecule of acetyl-$CoA$ at the input. In such a way, the continuous process of functioning of the Krebs cycle is running. This process has the autooscillatory character.

The studies of the model with the help of computational experiments showed that if system's parameters vary, the appearance of autooscillations with various frequencies, as well as chaotic oscillations, becomes possible. Oscillations with the same frequency will occur in all components of the given metabolic process. In the present work, we will study the dependence of autooscillations of the system on the parameter $k_8$, which determines the level of formation of $CO_2$ in the cycle of tricarboxylic acid.

The different types of obtained autooscillatory modes are studied with the help of the construction of phase-parametric diagrams. The abscissa axis shows the values of parameter $k_8$, and the axis of ordinates gives the values of chosen variable $E_1(t)$, for example. Moreover, we used the method of cutting. In the phase space of trajectories of the system, we place the cutting plane $S_2 = 0.8$. Such a choice is explained by the symmetry of oscillations of acetate relative to this plane in a lot of earlier calculated modes. For every given value of $k_8$, we observe the intersection of this plane by the trajectory in a single direction, when it approaches the attractor. The value of $E_1(t)$ is put onto the phase-parametric diagram. In the case where a multiple periodic limiting cycle arises, a number of points can be observed on the plane, and they will be the same in the period. If the deterministic chaos arises, the points of the intersection of the plane by the oscillating trajectory will be positioned chaotically.

In Fig. 2,a-d, we show the phase-parametric diagrams for the variable $E_1(t)$ versus the parameter $k_8$ changing in the appropriate intervals.

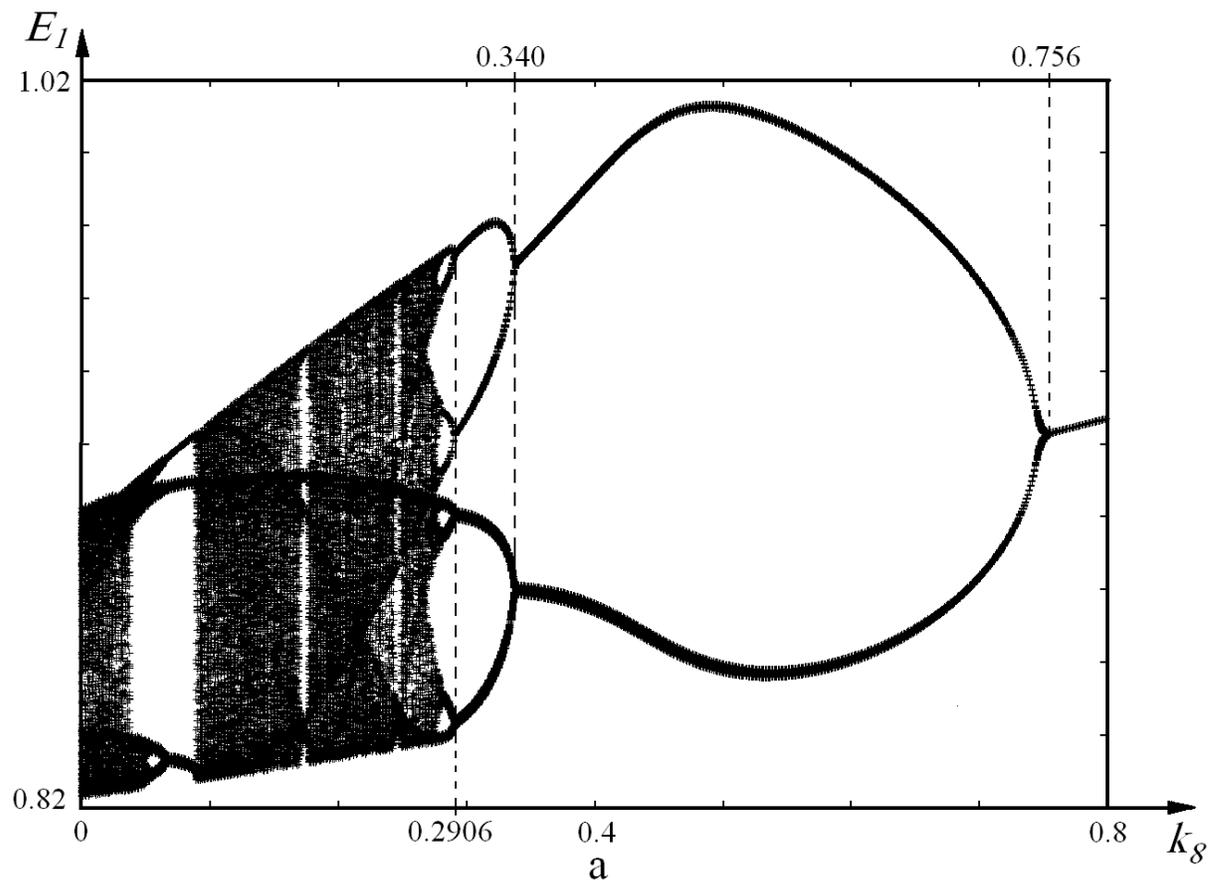

a

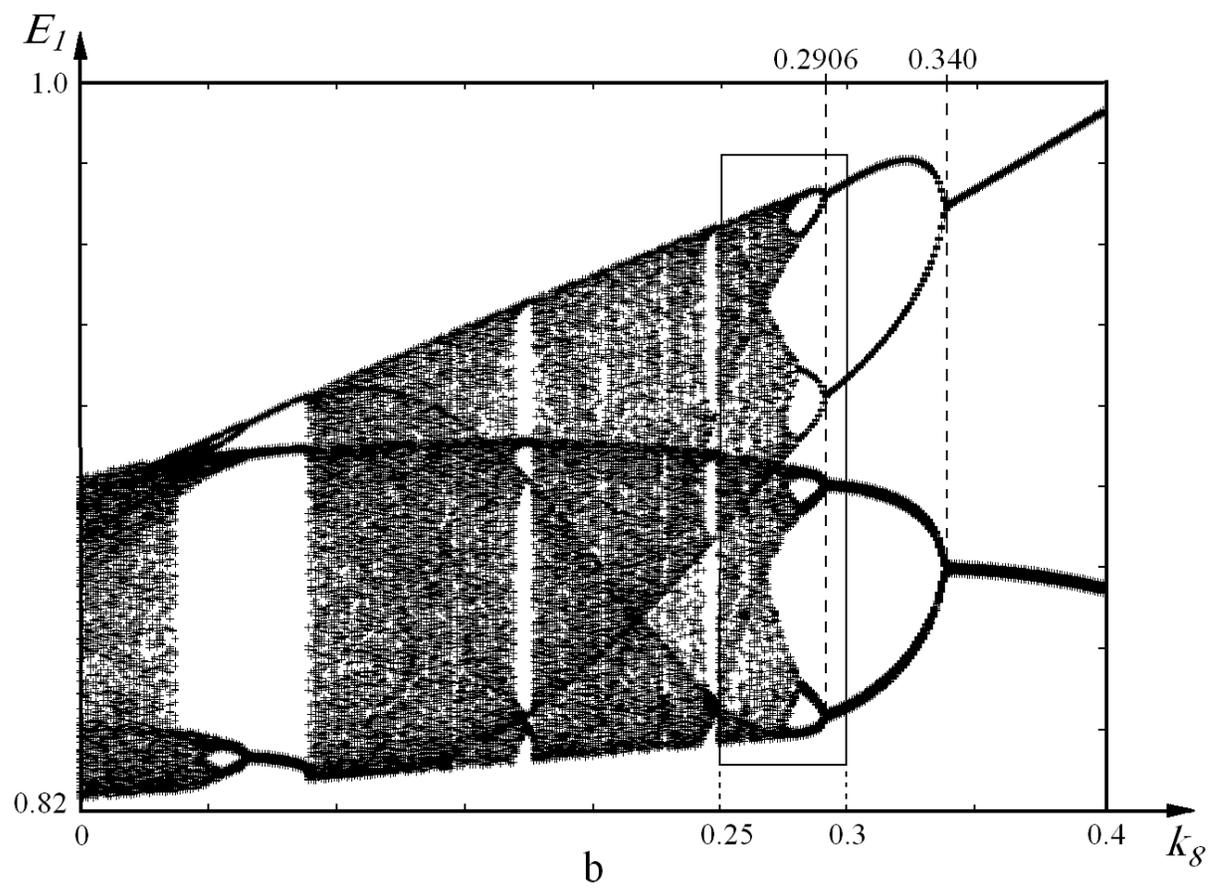

b

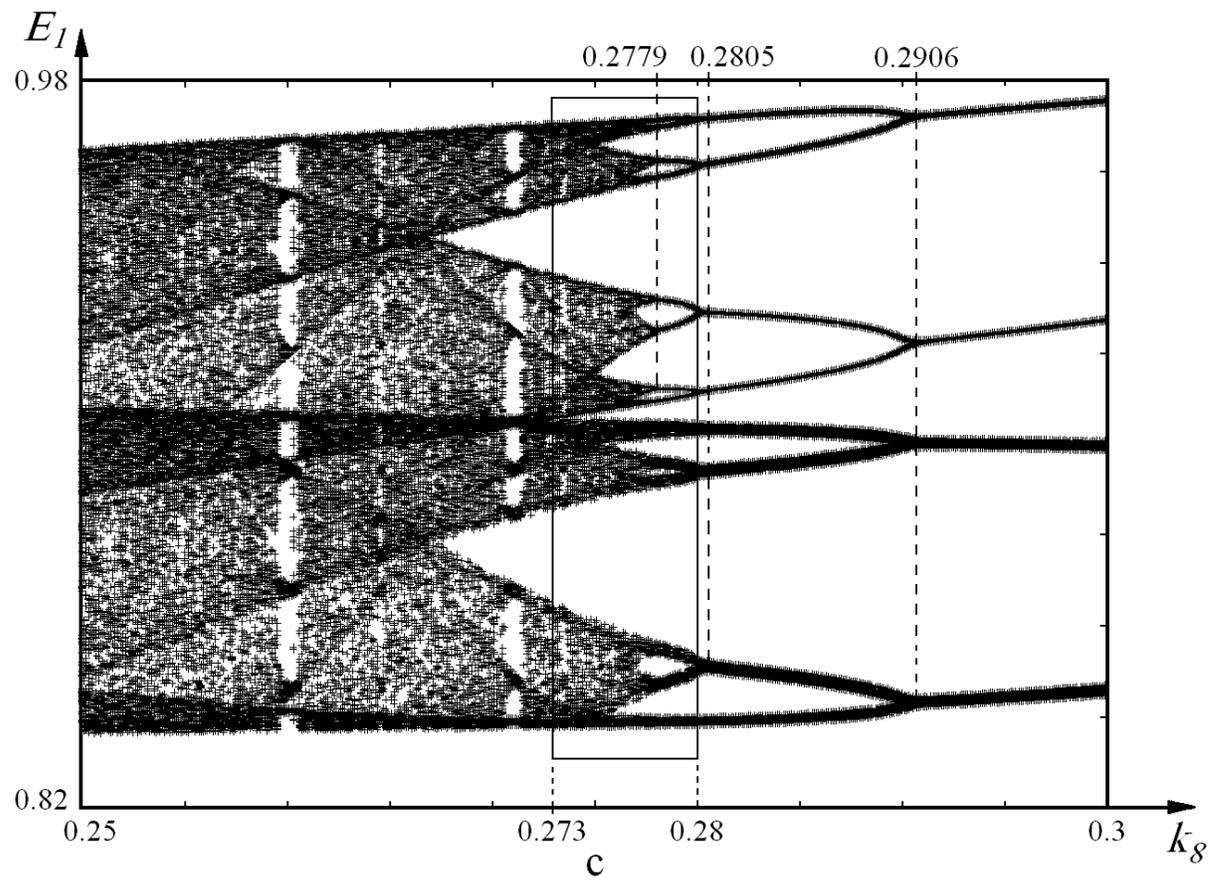

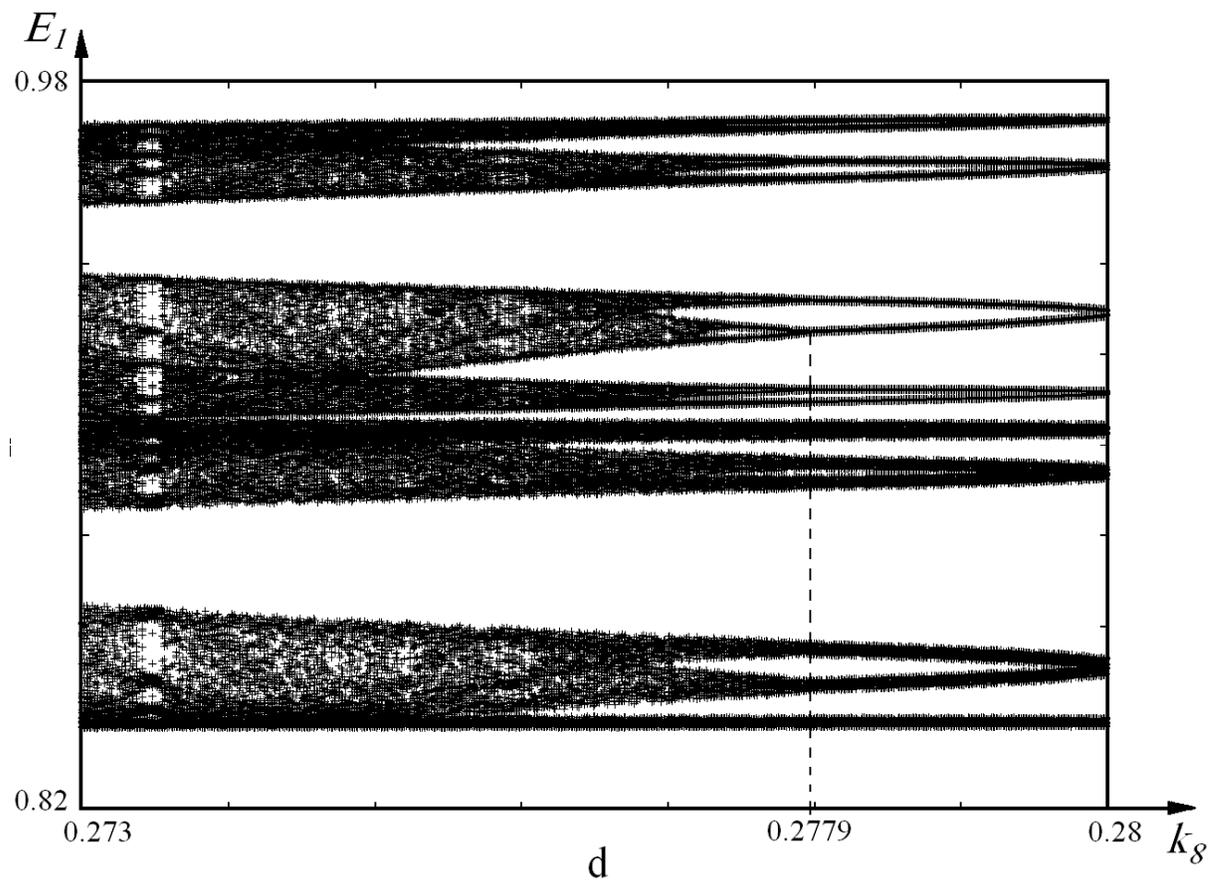

Fig. 2. Phase-parametric diagram for the variable $E_1(t)$:

a - $k_8 \in (0., 0.8)$; b - $k_8 \in (0., 0.4)$; c - $k_8 \in (0.25, 0.3)$; d - $k_8 \in (0.273, 0.28)$.

As the parameter $k_8$ decreases, there occurs the subsequent doubling of the multilicity of the autoperiodic process. Such a sequence of the appearance of bifurcations creates a cascade of bifurcations, namely the Feigenbaum sequence [44]. After the multiple doubling of a period, the modes of aperiodic oscillations are eventually observed in the system. In other words, a chaos arises. As the parameter $k_8$ decreases further, we see the appearance of the windows of periodicity on the phase-parametric diagrams. The deterministic chaos is destroyed, and the periodic and quasiperiodic modes are established. The trajectory of a strange attractor in the chaotic mode is tightened to a regular attractor of the autoperiodic mode. We observe the self-organization in the system. Then the windows of periodicity are destroyed, and the chaotic modes arise again. Moreover, the transitions "order—chaos" and "chaos—order" happen. There occurs the adaptation of the metabolic process to varying conditions.

It is seen from the presented figures that, as the scale decreases, every subsequent phase-parametric diagram with doubling of a cycle and its windows of periodicity are identicat to those of the previous diagram, as the scale decreases. The given sequence of bifurcations has a self-similar fractal structure.

In Figs. 3,e-f and 4, we present the examples of the projections of phase portraits for some values of parameter $k_8$, according to the phase-parametric diagram in Fig. 2.

In Fig. 5, we show the constructed kinetic curves for a strange attractor formed at $k_8 = 0.12$.

These figures indicate a variation of the dynamics of a metabolic process of the Krebs cycle, which depends on the intensity of formation of the final oxidation product, $CO_2$.

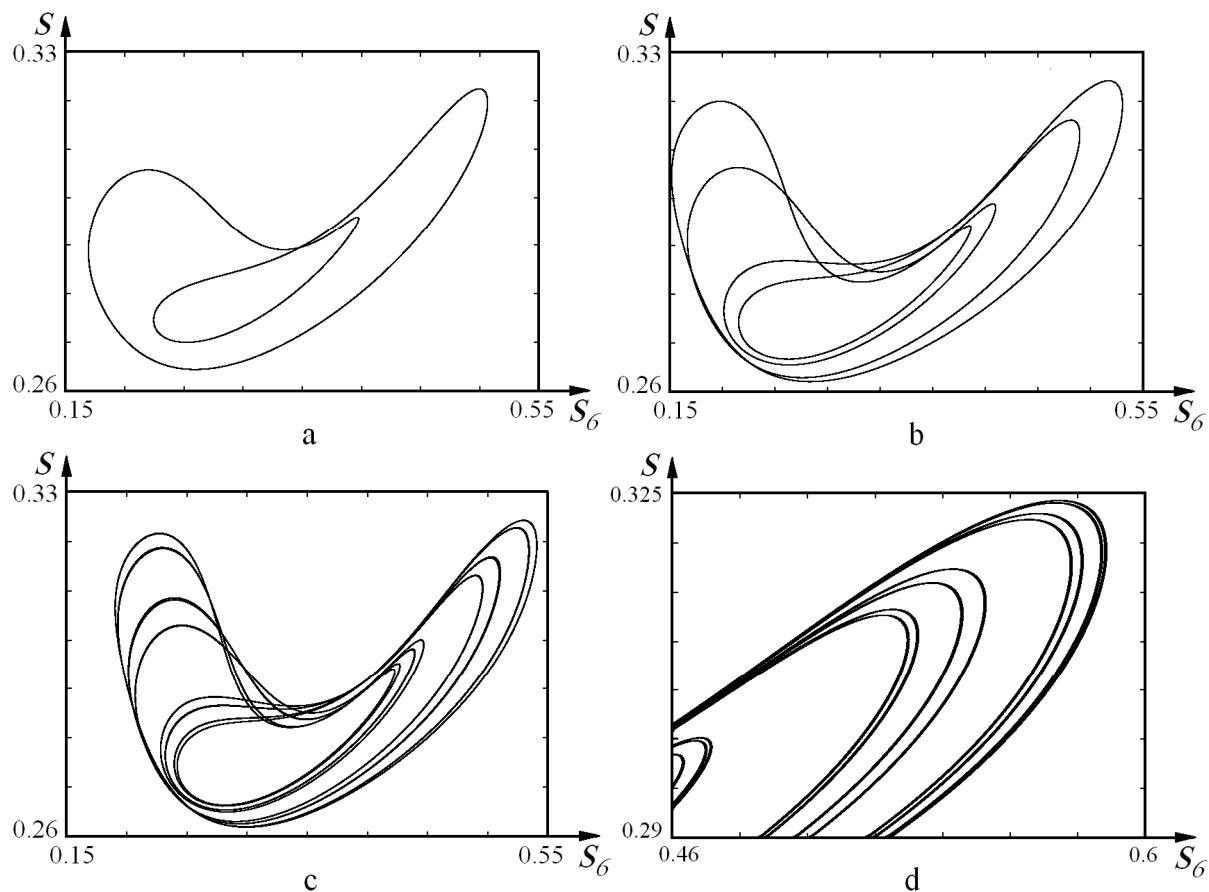

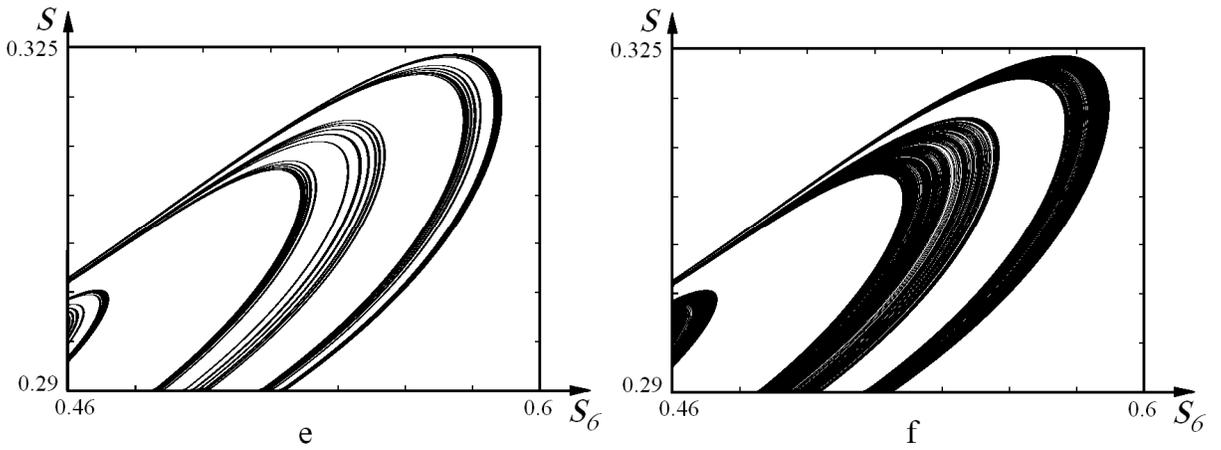

Fig. 3. Projections of system's phase portraits:
a – regular attractor $2 \cdot 2^1$, $k_8 = 0.5$; b – regular attractor $2 \cdot 2^2$, $k_8 = 0.3$;
c - regular attractor $2 \cdot 2^4$, $k_8 = 0.28$; d - regular attractor $2 \cdot 2^8$, $k_8 = 0.278$;
e - regular attractor $2 \cdot 2^{16}$, $k_8 = 0.277$; f - strange attractor $2 \cdot 2^x$, $k_8 = 0.275$.

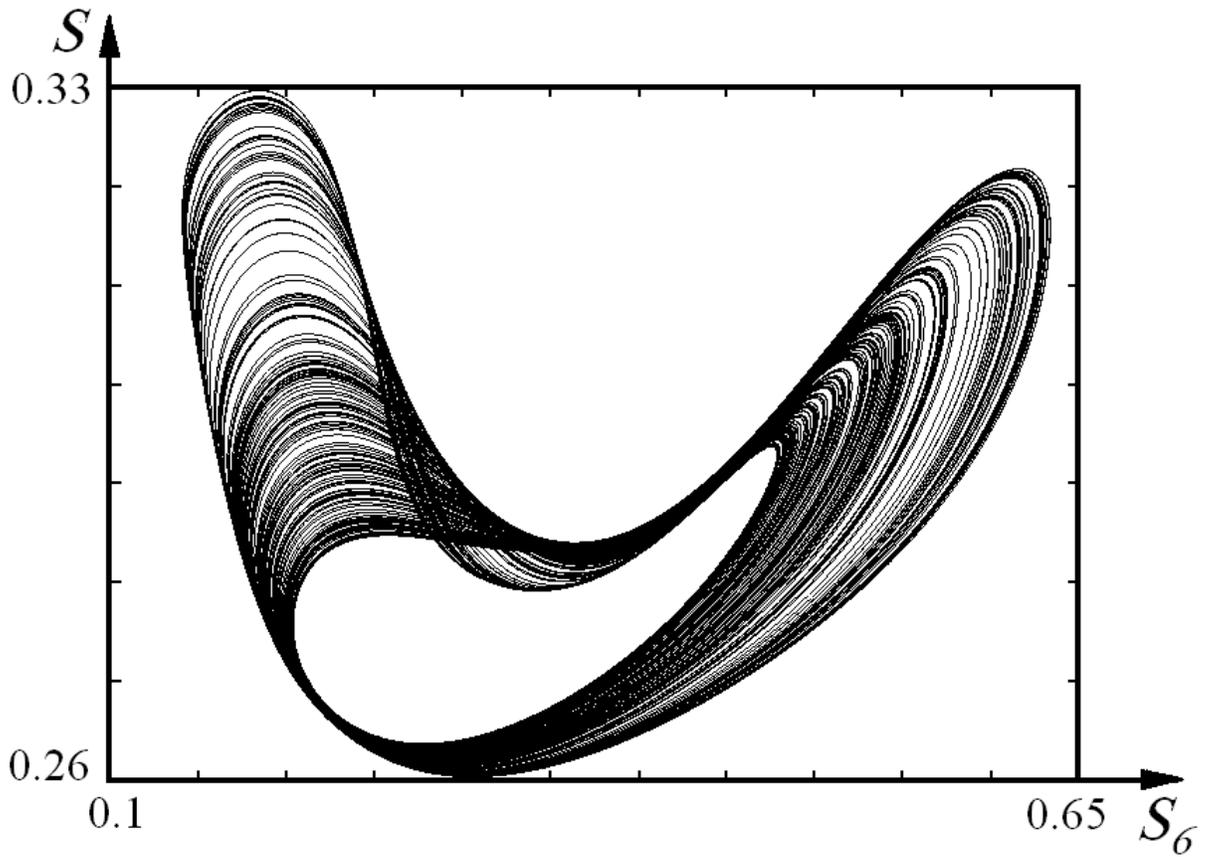

Fig. 4. Projection of the phase portrait of the strange attractor $2 \cdot 2^x$ for $k_8 = 0.12$.

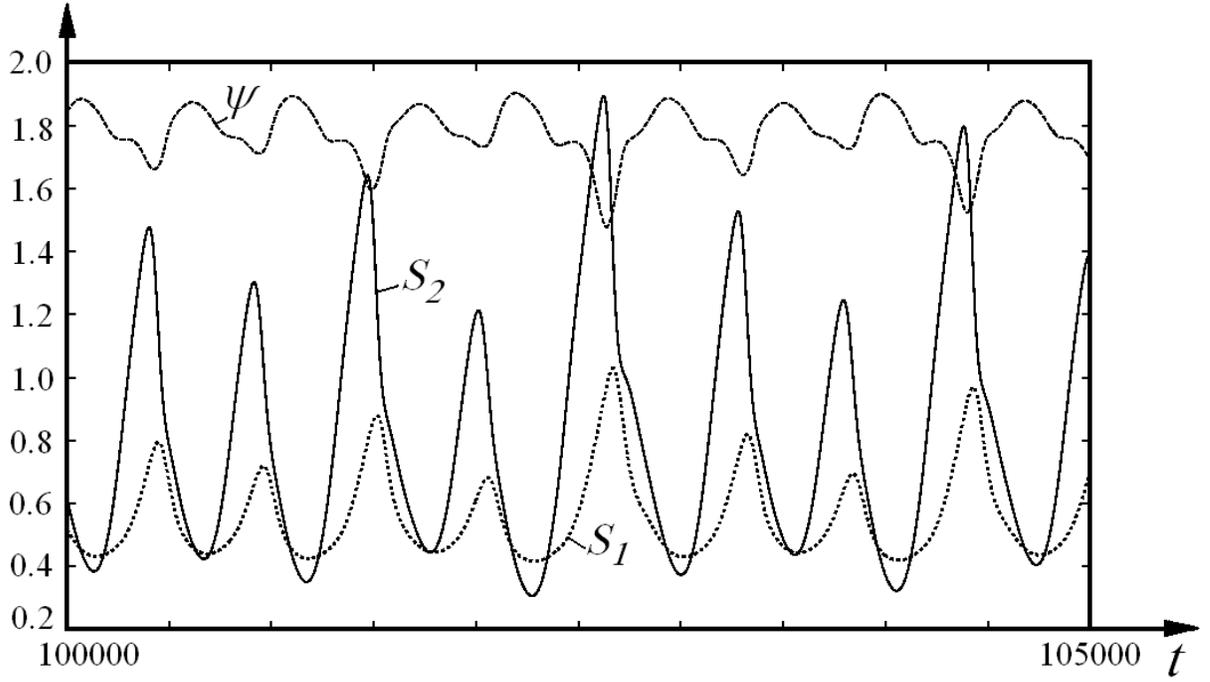

Fig. 5. Kinetic curve for the components $S_1$, $S_2$, and $\psi$ of the Krebs cycle in the mode of the strange attractor $2 \cdot 2^x$ for $k_8 = 0.12$.

In order to uniquely identify the type of obrained attractors and to determine their stability, we calculated the full spectra of Lyapunov indices and their sum $\Lambda = \sum_{j=1}^{10} \lambda_j$ for the chosen points. The calculation was carried out by Benettin's algorithm with the orthogonalization of the vectors of perturbations by the Gram--Schmidt method [26].

The algorithm of calculations of the full spectrum of Lyapunov indices is as follows. Taking some point on the attractor $\overline{X_0}$ as the initial one, we trace the evolution of $N$ vectors of perturbations along with the trajectory leafing this point. In our case, $N = 19$ is the number of variables of the system. We have numerically solved the initial equations describing the system, which are supplemented by 19 collections of equations in variations. As the initial vectors of perturbations, we take a collection of vectors $\overline{b_1^0}$, $\overline{b_2^0}, \ldots \overline{b_{19}^0}$, which are orthogonal to one another and are normalized to 1. In some time $T$, the trajectory comes to a point $\overline{X_1}$, and the vectors of perturbations will be $\overline{b_1^1}$, $\overline{b_2^1}$, $\ldots \overline{b_{19}^1}$. Their renormalization and orthogonalization by the Gram--Schmidt method are performed by the following scheme:

$$\overline{b_1^1} = \frac{\overline{b_1}}{\|\overline{b_1}\|},$$

$$\overline{b_2'} = \overline{b_2^0} - (\overline{b_2^0}, \overline{b_1^1})\overline{b_1^1}, \quad \overline{b_2^1} = \frac{\overline{b_2'}}{\|\overline{b_2'}\|},$$

$$\overline{b_3'} = \overline{b_3^0} - (\overline{b_3^0}, \overline{b_1^1})\overline{b_1^1} - (\overline{b_3^0}, \overline{b_2^1})\overline{b_2^1}, \quad \overline{b_3^1} = \frac{\overline{b_3'}}{\|\overline{b_3'}\|},$$

$$\overline{b'_4} = \overline{b_4^0} - (\overline{b_4^0}, \overline{b_1^1})\overline{b_1^1} - (\overline{b_4^0}, \overline{b_2^1})\overline{b_2^1} - (\overline{b_4}, \overline{b_3^1})\overline{b_3^1}, \quad \overline{b_4^1} = \frac{\overline{b'_4}}{\|b'_4\|},$$

..................................................................................................

$$\overline{b'_{19}} = \overline{b_{19}^0} - (\overline{b_{19}^0}, \overline{b_1^1})\overline{b_1^1} - (\overline{b_{19}^0}, \overline{b_2^1})\overline{b_2^1} - (\overline{b_{19}}, \overline{b_3^1})\overline{b_3^1} - ... - (\overline{b_{19}}, \overline{b_{18}^1})\overline{b_{18}^1}, \quad \overline{b_{19}^1} = \frac{\overline{b'_{19}}}{\|b'_{19}\|},$$

Then we continued the calculations, by starting from the point $\overline{X_1}$ and the vectors of perturbations $\overline{b_1^1}$, $\overline{b_2^1}$, ... $\overline{b_{19}^1}$. In the following time interval $T$, we have a new collection of the vectors of perturbations $\overline{b_1^2}$, $\overline{b_2^2}$, ... $\overline{b_{19}^2}$, which undergoes again the renormalization and orthogonalization by the above-indicated scheme. The described sequence of actions is repeated a sufficiently large number M of times. In this case as a result of calculations, we determine the values of sums

$$S_1 = \sum_{i=1}^{M} \ln\|b_1'^i\|, \quad S_2 = \sum_{i=1}^{M} \ln\|b_2'^i\|, ... S_{19} = \sum_{i=1}^{M} \ln\|b_{19}'^i\|,$$

where the vectors of perturbations are present prior to the renormalization, but after the orthogonalization.

We estimated 19 Lyapunov indices in the following way:

$$\lambda_j = \frac{S_j}{MT}, \quad i = 1, 2, ...19.$$

The calculation of Lyapunov indices from this multidimensional system on a personal computer meets certain difficulties. The mathematical model of the given biochemical system contains many variables and parameters. The limitations in the solution of such problems arise due to the insufficient random-access memory of a computer in the processing of the $n \times n$ matrix of small perturbations. In addition, any inaccuracy on the stage of programming will essentially affect the redefinition of the vectors of perturbations, their orthogonalization, and, as a consequence, the result of calculations. Nevertheless, we solved the problem and obtained certain results. Below for the sake of comparison, we present the spectra of Lyapunov indices for some modes of the system. For brevity without any loss of information, we give the values of indices up to the fourth decimal point.

The ratios of the values of Lyapunov indices $\lambda_1 > \lambda_2 > \lambda_3 > ... > \lambda_{19}$ serve as the criterion of the validity of calculations. For a regular attractor, we have obligatorily $\lambda_1 \approx 0$. The remaining indices can be also $\approx 0$ in some cases. In some other cases, they are negative. The zero value of the first Lyapunov index testifies to the presence of a stable limiting cycle.

For a strange attractor, at least one Lyapunov index must be positive. After it, the zero index follows. The next indices are negative. The presence of negative indices means the contraction of system's phase space in the corresponding directions, whereas the positive indices indicate the dispersion of trajectories. Therefore, there occurs the mixing of trajectories in narrow places of the phase space of the system, i.e., there appears the deterministic chaos. The Lyapunov indices contain obligatorily the zero index, which means the conservation of the aperiodic trajectory of an attractor in some region of the phase space and the existence of a strange attractor.

For $k_8 = 0.01$, the strange attractor $2 \cdot 2^x$ arises.

We have $\lambda_1 - \lambda_{19}$: .0007; .0000; -.0040; -.0125; -.0196; -.0200; -.0290; -.0299; -.0317; -.0416; -.0416;
-.0416; -.0458; -.0816; -.0874; -.0874; -.1181; -.1539; -.2222; $\Lambda$ = -1.0672.

For $k_8 = 0.075$ – regular attractor $3 \cdot 2^0$ (see the window of periodicity in Fig. 2,b).

$\lambda_1 - \lambda_{19}$: .0000; -.0004; -.0040; -.0117; -.0194; -.0211; -.0285; -.0285; -.0326; -.0406; -.0406; -.0406; -.0451; -.0819; -.0883; -.0883; -.1182; -.1563; -.2241; $\Lambda$ = -1.0702.

For $k_8 = 0.12$ – strange attractor $2 \cdot 2^x$;

$\lambda_1 - \lambda_{19}$: .0008; .0000; -.0040; -.0125; -.0192; -.0210; -.0287; -.0300; -.0324; -.0406; -.0406; -.0406; -.0457; -.0822; -.0879; -.0879; -.1172; -.1542; -.2212; $\Lambda$ = -1.0653.

For $k_8 = 0.27$ – strange attractor $2 \cdot 2^x$;

$\lambda_1 - \lambda_{19}$: .0002; .0000; -.0040; -.0125; -.0192; -.0219; -.0281; -.0310; -.0320; -.0387; -.0387; -.0387; -.0436; -.0842; -.0889; -.0889; -.1185; -.1541; -.2209; $\Lambda$ = -1.0638.

For $k_8 = 0.275$ – strange attractor $2 \cdot 2^x$;

$\lambda_1 - \lambda_{19}$: .0001; .0000; -.0040; -.0125; -.0192; -.0219; -.0280; -.0312; -.0323; -.0383; -.0383; -.0383; -.0434; -.0842; -.0890; -.0890; -.1184; -.1540; -.2212; $\Lambda$ = -1.0631.

For $k_8 = 0.278$ – regular attractor $2 \cdot 2^8$;

$\lambda_1 - \lambda_{19}$: .0000; .0000; -.0041; -.0123; -.0193; -.0219; -.0283; -.0308; -.0320; -.0384; -.0384; -.0384; -.0435; -.0842; -.0890; -.0890; -.1187; -.1538; -.2212; $\Lambda$ = -1.0634.

For $k_8 = 0.278$ – regular attractor $2 \cdot 2^8$;

$\lambda_1 - \lambda_{19}$: .0000; .0000; -.0041; -.0123; -.0193; -.0219; -.0283; -.0308; -.0320; -.0384; -.0384; -.0384; -.0435; -.0842; -.0890; -.0890; -.1187; -.1538; -.2212; $\Lambda$ = -1.0634.

For $k_8 = 0.28$ – regular attractor $2 \cdot 2^4$;

$\lambda_1 - \lambda_{19}$: .0000; .0000; -.0041; -.0123; -.0193; -.0219; -.0282; -.0307; -.0322; -.0384; -.0384; -.0384; -.0436; -.0842; -.0890; -.0890; -.1186; -.1540; -.2212; $\Lambda$ = -1.0633.

The presented results of calculations indicate that the sum $\Lambda$ of all indices, which determine the flow divergencies and, hence, the evolution of the phase volume along the trajectory, is maximal for the regular attractor $3 \cdot 2^0$. It arises in the window of periodicity for $k_8 = 0.075$ ($\Lambda$ = -1.0702). For the strange attractors on the left and on the right (for $k_8 = 0.01$ and $k_8 = 0.12$), the divergencies are, respectively, $\Lambda$ = -1.0672 and $\Lambda$ = -1.0653. This means that the phase volume element for the given attractor is contracted, on the whole, stronger along the trajectory. Here, we observe the self-organization of a stable cycle from chaotic modes. The Krebs cycle is adapted to the varying conditions.

By the given Lyapunov indices for strange attractors, we determine the KS-entropy (the Kolmogorov--Sinai entropy) [45]. By the Pesin theorem [46], the KS-entropy $h$ corresponds the sum of all positive Lyapunov characteristic indices:

The KS-entropy allows us to judge about the rate, with which the information about the initial state of the system is lost. The positivity of the given entropy is a criterion of the chaos. This gives possibility to qualitatively estimate the properties of attractor's local stability.

We determine also the quantity inverse to the KS-entropy, $t_{min}$. This is the time of a mixing in the system. It characterizes the rate, with which the initial conditions will be forgotten. For $t \ll t_{min}$, the behavior of the system can be predicted with sufficient accuracy. For $t > t_{min}$, only a probabilistic description is possible. The chaotic mode is not predictable due to the loss of the memory of initial conditions. The quantity $t_{min}$ is called the Lyapunov index and characterizes the "predictability horizon" of a strange attractor.

In order to classify the geometric structure of strange attractors, we calculated the dimension of their fractality. The strange attractors are fractal sets and have the fractional Hausdorff-Besicovitch dimension. But its direct calculation is a very labor-consuming task possessing no standard algorithm. Therefore, as a quantitative measure of the fractality, we calculated the Lyapunov dimension of attractors by the Kaplan--Yorke formula [47, 48]:

$$D_{F_r} = m + \frac{\sum_{i=1}^{m} \lambda_i}{|\lambda_{m+1}|}, \qquad (11)$$

where $m$ is the number of the first Lyapunov indices ordered by their decreasing. Their sum $\sum_{i=1}^{m} \lambda_i \geq 0$, and $m+1$ is the number of the first Lyapunov index, whose value $\lambda_{m+1} < 0$.

For the above-considered strange attractors $2^\infty$, we obtained the following indices.
For $k_8 = 0.01$: $h = 0.0007$, $t_{min} = 1428.6$, $D_{F_r} = 2.175$.
For $k_8 = 0.12$: $h = 0.0008$, $t_{min} = 1250$, $D_{F_r} = 2.2$.
For $k_8 = 0.27$: $h = 0.0002$, $t_{min} = 5000$, $D_{F_r} = 2.05$.
For $k_8 = 0.275$: $h = 0.0001$, $t_{min} = 10000$, $D_{F_r} = .2.025$.

By these indices, we can judge about the difference of the given strange attractors.

## CONCLUSIONS

With the help of the mathematical model of the Krebs cycle, we have studied the dependence of the cyclicity of the metabolic process on the amount of a final product of the oxidation, i.e., on the amount of the formed carbon dioxide. The multiplicity of the cycle is doubled by the Feigenbaum scenario, until the aperiodic modes of strange attractors arise. From them as a result of the self-organization, the stable periodic modes appear. This means that the system is adapted to the varying conditions. We have calculated the full spectra of Lyapunov indices and the divergencies for various modes. For the strange attractors, we have determined the KS-entropies, "predictability horizons," and Lyapunov dimensions of attractors. The results obtained allow us to study the structural-functional connections of the cycle of tricarboxylic acids, their influence on the cyclicity of metabolic oscillations in a cell, and the physical laws of self-organization in it.

*The work is supported by the project No. 0113U001093 of the National Academy of Sciences of Ukraine.*

**References**

1. H. A. Krebs and W. A. Johnson. The role of citric acid in intermediate metabolism in animal tissues. *Enzymologia*, 4, 148–156, 1937.
2. R. Bohnensack and E .E. Sel'kov. Stoichiometric regulation in the citric acid cycle. II. Non-linear interactions. *Studia biophysica*, 66, 47–63, 1977.
3. A. E. Lyubarev and B. I. Kurganov. Supermolecular organization of enzymes of the cycle of tricarboxylic acids. *Molek. Biol.*, 21, 5, 1286–1296, 1987.
4. M. N. Kondrashova. Structural kinetic organization of Supermolecular organization of enzymes of the cycle of tricarboxylic acids under the active functioning of mitochondria. *Biofiz.*, 34, 3, 450–457, 1989.
5. E. M. T. El-Mansi, G. C. Dawson and C. F. A. Bryce. Steady-state modelling of metabolic flux between the tricarboxylic acid cycle and the glyoxylate bypass in Escherichia coli. *Comput. Applic. Biosci.*, 10, 3, 295–299, 1994.
6. R. Ramakrishna, J.S. Edwards, A. McCulloch and B. O. Palsson. Flux-balance analysis of mitochondrial energy metabolism: consequences of systemic stoichiometric constraints. *Am. J. Physiol. Regul. Integr. Comp. Physiol.*, 280, 3, R695–R704, 2001.
7. S. Cortassa, M. A. Aon, E. Marban, R. L. Winslow and B. O'Rourke. An integrated model of cardiac mitochondrial energy metabolism and calcium dynamics. *Biophys. J.*, 84, 2734–2755, 2003.

8. K. Yugi and M. Tomita. A general computational model of mitochondrial metabolism in a whole organelle scale. *Bioinformatics*, 20, 1795–1796, 2004.
9. V. K. Singh and I. Ghosh. Kinetic modeling of tricarboxylic acid cycle and glyoxylate bypass in Mycobacterium tuberculosis, and its application to assessment of drug targets. *Theor. Biol. and Med. Model.*, 3, 27, 2006.
10. E. Mogilevskaya, O. Demin and I. Goryanin. Kinetic model of mitochondrial Krebs cycle: unraveling the mchanism of salicylate hepatotoxic effects. *J. of Biol. Phys.*, 32, 3-4, 245–271, 2006.
11. V. P. Gachok. *Kinetics of Biochemical Processes*, Naukova Dumka, Kiev, 1988 (in Russian).
12. V. P. Gachok. *Strange Attractors in Biosystems*, Naukova Dumka, Kiev, 1989 (in Russian).
13. J. Monod. *Recherches sur la Croissanse des Cultures Bacteriennes*, Hermann, Paris, 1942.
14. V. S. Podgorskii. *Physiology and Metabolism of Methanol-Consuming Yeast*, Naukova Dumka, Kiev, 1982 (in Russian).
15. L. N. Drozdov-Tikhomirov and N. T. Rakhimova. On the maximum yoeld of the biomass of methanol-consuming yeast. *Mikrobiol.*, 55, 5, 775–780, 1986.
16. G. Yu. Riznichenko. *Mathematical Models in Biophysics and Ecology*, Inst. of Comput. Studies, Moscow-Izhevsk, 2003 (in Russian).
17. C. M. Watteeuw, W. B. Armiger, D. L. Ristroph and A. E. Humphrey. Production of single cell protein from ethanol by fed-batch process. *Biotechnol. and Bioeng.*, 21, 5, 1221–1237, 1979.
18. W. B. Armiger, A. R. Moreira, J. A. Phillips and A. E. Humphrey. Modeling cellulose digestion for single cell protein. *Utilization of Cellulose Materials in Inconventional Food Production*, Plenum Press, New York, 1979, 111–117.
19. V. I. Grytsay and I. V. Musatenko. Self-organization and fractality in a metabolic processes of the Krebs cycle. *Ukr. Biokh. Zh.*, 85, 5, 191–199, 2013.
20. E. E. Selkov. Self-oscillations in glycolysis. *Europ. J. Biochem.*, 4, 79–86, 1968.
21. B. Hess and A. Boiteux. Oscillatory phenomena in biochemistry. *Annu. Rev. Biochem.*, 40, 237–258, 1971.
22. A. Goldbeter and R. Lefer. Dissipative structures for an allosteric model. Application to glycolytic oscillations. *Biophys J.*, 12, 1302–1315, 1972.
23. A. Godlbeter and R. Caplan. Oscillatory enzymes. *Annu. Rev. Biophys. Bioeng.*, 5, 449–476, 1976.
24. *Chaos in Chemical and Biochemical Systems*, edited by R. Field, L. Györgyi, World Scientific, Singapore, 1993.
25. V. S. Anishchenko. *Complex Oscillations in Simple Systems*, Nauka, Moscow, 1990 (in Russian).
26. S. P. Kuznetsov. *Dynamical Chaos*, Nauka, Moscow, 2001 (in Russian).
27. V. P. Gachok and V. I. Grytsay. Kinetic model of macroporous granule with the regulation of biochemical processes. *Dokl. Akad. Nauk SSSR*, 282, 51–53, 1985.
28. V. P. Gachok, V. I. Grytsay, A. Yu. Arinbasarova, A. G. Medentsev, K. A. Koshcheyenko and V. K. Akimenko. Kinetic model of hydrocortisone 1-en dehydrogenation by Arthrobacter globiformis. *Biotechn. Bioengin.*, 33, 661–667, 1989.
29. V. P. Gachok, V. I. Grytsay, A. Yu. Arinbasarova, A. G. Medentsev, K. A. Koshcheyenko and V. K. Akimenko. A kinetic model for regulation of redox reactions in steroid transformation by Arthrobacter globiformis cells. *Biotechn. Bioengin.*, 33, 668–680, 1989.
30. V. I. Grytsay. Self-organization in the macroporous structure of the gel with immobilized cells. Kinetic model of bioselective membrane of biosensor. *Dopov. Nats. Akad. Nauk Ukr.*, 2, 175–179, 2000.
31. V. I. Grytsay. Self-organization in a reaction-diffusion porous media. *Dopov. Nats. Akad. Nauk Ukr.*, 3, 201–206, 2000.
32. V. I. Grytsay. Ordered structure in a mathematical model biosensor. *Dopov. Nats. Akad. Nauk Ukr.*, 11, 112–116, 2000.
33. V.I. Grytsay. Self-organization of biochemical process of immobilized cells of bioselective membrane biosensor. *Ukr. J. Phys.*, 46, 1, 124–127, 2001.

34. V. V. Andreev and V. I. Grytsay. The modeling of nonactive zones in porous granules of a catalyst and in a biosensor. *Matem. Modelir.*, 17, 2, 57–64, 2005.
35. V. V. Andreev and V. I. Grytsay. Influence of heterogeneity of diffusion-reaction process for the formation of structures in the porous medium. *Matem. Modelir.*, 17, 6, 3–12, 2005.
36. V. I. Grytsay and V. V. Andreev. The role of diffusion in the active structures formation in porous reaction-diffusion media. *Matem. Modelir.*, 18, 12, 88–94, 2006.
37. V. I. Grytsay. Unsteady conditions in porous reaction-diffusion. *Medium. Romanian J. Biophys.*, 17, 1, 55–62, 2007.
38. V. I. Grytsay. The uncertainty in the evolution structure of reaction-diffusion medium bioreactor. *Biofiz. Visn.*, 2, 92–97, 2007
39. V. I. Grytsay. Formation and stability of morphogenetic fields of immobilized cell in bioreactor. *Biofiz. Visn.*, 2, 25–34, 2008.
40. V. I. Grytsay. Structural instability of a biochemical process. *Ukr. J. Phys.,* 55, 5, 599–606, 2010.
41. V. I. Grytsay and I. V. Musatenko. Self-oscillatory dynamics of the metabolic process in a cell. *Ukr. Biochem. J.*, 85, 2, 93–104, 2013.
42. V. I. Grytsay and I. V. Musatenko. The structure of a chaos of strange attractors within a mathematical model of the metabolism of a cell. *Ukr. J. of Phys.*, 58, 7, 677–686, 2013.
43. V. I. Grytsay and I. V. Musatenko. A mathematical model of the metabolism of a cell. Self-organisation and chaos. Chaotic modeling and simulation. *CMSIM*, 4, 539-552, 2013.
44. M. J. Feigenbaum. Quantative universality for a class of nonlinear transformations. *J. Stat. Phys.*, 19, 1, 25–52, 1978.
45. A. N. Kolmogorov. On the entropy per unit time as a metric invariant of automorphisms. *DAN SSSR*, 154, 754–755, 1959.
46. Ya. B. Pesin. Characteristic Lyapunov indices and the ergodic theory. *Usp. Mat. Nauk*, 32, 4, 55–112, 1977.
47. J. L. Kaplan and J. A. Yorke. The onset of chaos in a fluid flow model of Lorenz. *Ann. N. Y. Acad. Sci.*, 316, 400–407, 1979.
48. J. L. Kaplan and J. A. Yorke. A chaotic behaviour of multidimensional differential equations. *Functional Differential Equations of Fixed Points*, edited by H. O. Peitgen, H. O. Walther, Springer, Berlin, 1979, 204-227.